# Oxygen Vacancy-Induced Monoclinic Dead Layers in Ferroelectric $Hf_xZr_{1-x}O_2$ With Metal Electrodes

Tanmoy Kumar Paul[1], Atanu Kumar Saha[1], and Sumeet Kumar Gupta[1]
[1]Purdue University, West Lafayette, Indiana, USA, email: paul115@purdue.edu

**In this work, we analyze dead layer comprising non-polar monoclinic (m) phase in $Hf_xZr_{1-x}O_2$ (HZO)-based ferroelectric (FE) material using first principles analysis. We show that with widely used tungsten (W) metal electrode, the spatial distribution of the oxygen vacancy across the cross-section plays a key role in dictating the favorability of m- phase formation at the metal-$HfO_2$ interface. The energetics are also impacted by the polarization direction as well as the depth of oxygen vacancy, i.e., position along the thickness. At the metal - $HfO_2$ interface, when polarization points towards the metal and vacancy forms at trigonally bonded O atomic site, both interfacial relaxation and m- phase formation can lead to dead layers. For vacancies at other oxygen atomic sites and polarization direction, dead layer is formed due to sole interfacial relaxation with polar phase. We also establish the relative favorability of the m-phase dead layer for different *Zr* concentrations (*x*=1 and *x* = 0.5) and metal electrodes. According to our analysis, 50% *Zr* doped $HfO_2$ exhibits less probability of m-phase dead layer formation compared to pure $HfO_2$. Moreover, with electrodes consisting of noble metal (Pt, Pd, Os, Ru, Rh), m-phase dead layer formation is less likely. Therefore, for these metals, dead layer forms mainly due to the interfacial relaxation with polar phase.**

## 1. Introduction

Developing a deep understanding of the interfacial dead layers in ferroelectric (FE) capacitors is crucial as many of the ferroelectric attributes are directly impacted by them [1]. In general, the dead layer is defined as the region close to the metal interface with attributes such as permittivity different from the bulk values [2]. In the context of ferroelectric (FE) thin film devices, dead layers refer to those layers which do not exhibit hysteretic polarization switching [3]. Hence, the presence of dead layers alters the electrostatics of the ferroelectric material via depolarization field [4]. Increase in the dead layer thickness degrades the charge screening, leading to an increase in the depolarization field in the ferroelectric. Consequently, the dead layer affects the remanent polarization, coercive field, multi-domain formation, permittivity and other important ferroelectric characteristics. Further, understanding the impact of dead layer is important for optimizing the devices for various applications. For example, thickness of dead layer affects the domain density inside the FE, which in turn, impacts the applicability of the devices in logic and memory applications [5]. In the context of ferroelectric capacitors or ferroelectric transistor-based memories, dead layers are generally considered detrimental as they can degrade the memory window and retention in FE based devices [6]. On the other hand, in ferroelectric tunnel junctions (FTJs), pinned dead layer with lower dielectric constant can enhance the tunneling electroresistance [7]. Due to far reaching effects of the dead layers on both the characteristics and applications of FE-based devices, their mechanisms of formation and other attributes have been subjects of great interest.

Dead layers can form due to both interfacial relaxation and non-polar phase formation through chemical interaction [8]. In conventional Perovskite FEs, previous experimental and theoretical works have described polarization pinning near the interface as the source of dead layer formation [9] [10] [11]. Thickness of dead layer in perovskite FEs has been measured using several techniques i.e., small-signal capacitance modeling and hysteresis analysis [12] [13]. In recent times, $HfO_2$-based FE materials have sparked great interest due to their compatibility with CMOS technology. $HfO_2$-based devices demonstrate ferroelectricity in ultra-thin samples and therefore, even thin dead layers can have a large impact on the device performance.

There have been several reports of dead layer formation in experiments with $HfO_2$-based FE devices. For example, interfacial oxygen vacancy-rich layer forms in HZO with TiN electrode which has been reported as a primary cause of reduced remnant polarization [14] [15]. According to a previous report, bottom electrode is more prone to dead layer formation compared to top electrode [16] in HZO. The thickness of dead layer depends on the electrode material, specifically their oxidation capacity, i.e., density of oxygen vacancy [17] [16]. Karbasian et al. showed that in $Hf_{0.8}Zr_{0.2}O_2$ thin films, capping layer of Tungsten (W) results in lower interface oxidation and superior remnant polarization (suggesting reduced dead layer thickness) compared to TiN capping layer [17]. On the contrary, Oh et al. showed that with both TiN and W, HZO exhibits thick dead layers, and the FE phase is suppressed for sub-5nm film thicknesses. But with Pt electrode, HZO exhibits excellent FE characteristics with very thin dead layers [18]. This work predicts dead layers to be formed because of non-polar phase formation in the presence of oxygen vacancy [18]. In general, dead layer formation is dependent on the process parameters, device topology, temperature, and many other factors. Besides, electric field cycling can also modulate the dead layer in HZO as predicted by experiments [19] [20]. Given a large diversity of experimental results indicating the properties of dead layers, it is important to understand the potential mechanisms of dead layer formation in HZO in a greater detail. However, theoretical analyses for HZO in this regard have so far been lacking.

In this work, we address this gap by using first principles density functional theory calculations in HZO to investigate the following aspects of dead layer formation:

➢ Dead layer formation via non-polar monoclinic (m) phase near the interface of metal electrode and ferroelectric orthorhombic phase of $HfO_2$.
➢ Relationship of monoclinic dead layer formation with spatial distribution of oxygen vacancy.
➢ Effect of zirconium doping on dead layer formation.
➢ Impact of metal electrode on dead layer formation (which, we show is consistent with experimental observations [18]).

## 2. Results and Discussions

Previous experimental and theoretical investigations have established that orthorhombic (o-) $Pca2_1$ phase is responsible for ferroelectricity in $HfO_2$ based materials [21] [22]. On the other hand, monoclinic (m-) $P2_1/c$ phase is one of the non-polar phases of $HfO_2$. $HfO_2$ with predominant m-phase is widely used in CMOS technology as a high k-dielectric material [23]. The unit cells of both o-phase and m-phase consist of 4 Hf atoms and 8 O atoms. The upward and downward polarized unit cells of o-phase and unit cell of m-phase are shown in Figure 1a with their atomic coordinates in Supplementary Figure S1. Along one of the lateral directions ($x$), both the phases can be divided into two layers: (i) centric layer where the O atoms lie in the central plane to form tetrahedral bonding with four Hf atoms and (ii) eccentric layer where the O atoms deviate from the central plane to bond with three Hf atoms and form trigonal geometry. Centric and eccentric layers in the o-phase are widely denoted as dielectric/spacer and ferroelectric/polar layers, respectively [24] [25]. On the other hand, both centric and eccentric layers in m-phase have zero net polarization and are dielectric in nature.

Typically, in $HfO_2$-based devices, O-vacancies (i.e, absence of an O atom at a lattice site) are formed when Hf atoms are partially oxidized by the metal [26]. The distribution of the O-vacancies is random in nature. Thus, O-vacancies can form at the centric or the eccentric layers with a similar likelihood. The energy cost of removing an O atom is defined as O-vacancy formation energy ($V_0$). In this work, we consider neutral O-vacancies, for which $V_0$ is calculated at centric and eccentric layers using the following equation:

$$V_0 = E_{vacancy} - E_{perfect} + \frac{\mu_{O_2}}{2} \quad (1)$$

Here, $E_{vacancy}$ and $E_{perfect}$ are the energies with and without an O-vacancy, respectively. $\mu_{O_2}$ is the chemical potential of an O-molecule. We calculate $V_0$ at bulk using a single unit cell and $V_0$ at metal-HZO interface using asymmetric slab model of metal-HZO stack [27] [28].

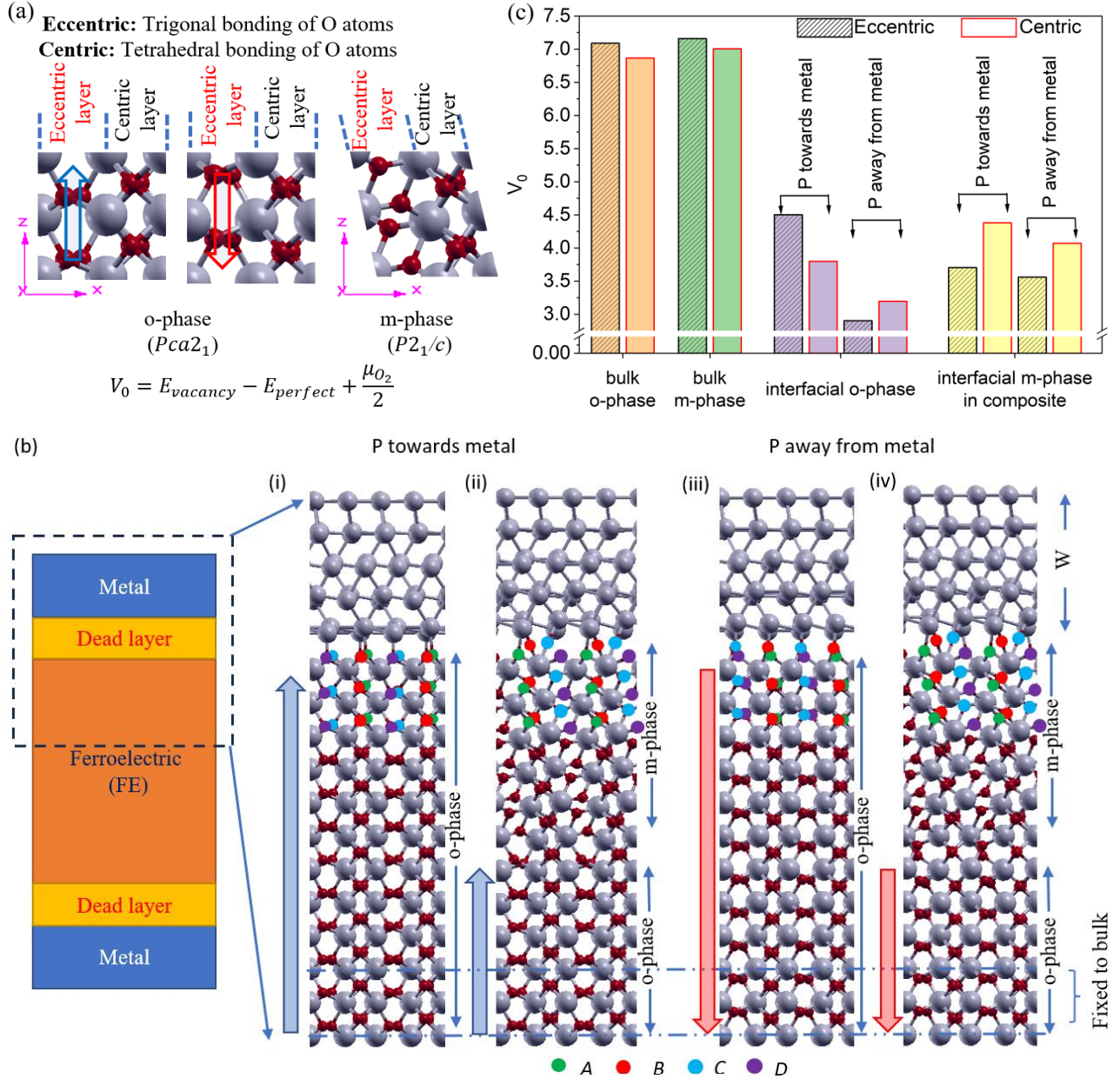

**Figure 1** (a) Polar orthorhombic (o) and non-polar monoclinic (m) phase of $HfO_2$. $x$- direction is segmented into centric and eccentric layers. (b) Atomic models of dead layer at metal (W)- FE $HfO_2$ interface. (i) pure o-phase and (ii) composite of m and o-phases when polarization directs towards the metal. (iii) pure o-phase and (iv) composite of m and o-phases when polarization directs away from the metal. Red colored *B* atom in centric layer has less O-vacancy formation energy than green colored *A* atom. Similarly, vacancy formation at the blue colored *C* atom in eccentric layer is energetically favorable than the purple *D* atom. Therefore, *A* and *C* atoms are considered for centric and eccentric O-vacancy, respectively. (c) Oxygen (O) vacancy formation energy ($V_0$) for centric and eccentric O atoms for bulk and interfacial o- phase and m-phase (1.5 nm thick). Bulk O-vacancy formation energy is much higher than interfacial O-vacancy.

As stated earlier, our goal is to analyze the possibility of m-phase dead layer formation at the interface of metal and FE o-phase. Therefore, the m-phases of different thicknesses are stacked between metal and o-phase keeping the total thickness of HfO$_2$ fixed. To mimic the bulk FE behavior away from the dead layer as depicted in Figure 1b, the unit cell of o-phase furthest from the interface is kept fixed at bulk atomic coordinates. Tungsten (W) is used as the metal electrode, unless otherwise specified. We refer to the structures with and without m-phase as the composite phase and pure o-phase, respectively. Now, for bulk o-HfO$_2$, $V_0$ does not depend on the direction of polarization due to the identical atomic bonding for up and down polarizations (refer to Figure 1a). Moreover, for both the bulk o- and m-HfO$_2$, all the eccentric O atoms in a unit cell have identical $V_0$. The same is true for all the centric O atoms. However, for the slab model used for the metal-HfO$_2$ interface, symmetry breaking near the interface results in the dependence of $V_0$ on the direction of polarization. In addition, the location of the vacancies plays a key role in dictating the O-vacancy formation energy.

To understand these dependencies, let us consider Figure 1b(i)-(iv), which illustrates the atomic models of the pure o-phases and composite phases with 1.5 nm m-phase for different polarization directions (i.e. polarization towards and away from the metal) in the FE domain. As the symmetries within the eccentric as well as the centric O atoms are broken at m-phase-metal interface, all the interfacial O atoms of the composite phase are energetically non-equivalent. We denote the centric O atoms by A and B, and eccentric O atoms by C and D, as depicted in Figure 1b(ii) and (iv). The O-vacancy formation energy for all the eccentric and centric O atoms of interfacial m-phase in the composite structure is shown in the bar plot of Supplementary Figure S2a. The plot suggests that vacancy formation at eccentric C atom is relatively easier compared to the eccentric *D* atom. In addition, vacancy formation at centric *B* atom is energetically favorable than centric *A* atom. For the composite phase, we, therefore, consider the sites of B and C atoms as the centric and eccentric O-vacancy, respectively for both polarization directions. Similarly, for pure o-phase, we consider the interfacial eccentric C atom or centric B atom (refer to Figure 1b(i) and (iii)) as the vacancy sites because they yield the lowest $V_0$ as shown in Supplementary Figure S2b. A comparative bar plot of thus considered $V_0$ for bulk and interfacial o/m-phases are shown in Figure 1c. To validate the results with metal-FE slab for a realistic device, we also perform similar analysis for metal-FE-metal (MFM) structure with 4.5 nm thick sample of HfO$_2$ with 1.5 nm m-phase dead layer at one of the interfaces. The atomic configurations and $V_0$ of pure o-phase and composite phases of MFM structures are depicted in Supplementary Figure S3. The plot validates that O-vacancy formation energies in MFM are consistent with those in metal-FE slab shown in Figure 1c. This suggests that the considered metal-FE slab model captures the energetics of a realistic MFM device.

Now, let us compare the bulk and interfacial O-vacancy formation energies in Figure 1c. Within bulk HfO$_2$, for both polarization directions of o-phase as well as m-phase, O-vacancy formation in the centric layer is energetically more favorable than the eccentric layer. This indicates a weaker bond strength in tetrahedral bonding compared to trigonal bonding which is consistent with bond lengths observed in previous works [29]. Moreover, the bulk O-vacancy formation energies are quite large - around 7 eV. Therefore, in the pristine state, O-vacancy formation within bulk HfO$_2$ is quite less likely. Bulk vacancies are created after bond weakening due to fatigue as observed during experiments [30] [31]. On the other hand, $V_0$ at interfacial o-phase and m-phase are much less, as depicted in Figure 1c. As a result, O vacancies are more favorable to form at the interfaces of HfO$_2$ and metal electrode. Therefore, we will first analyze the effect of O-vacancies on m-phase dead layer formation at the metal-FE interface and later discuss the effect of shifting of O vacancies away from the interface.

### 2.1 Oxygen Vacancies at the Interface

### 2.1.1 Polarization pointing towards metal

Let us first analyze the metal interface with polarization pointing towards it. We calculate the energies of the pure o-phase and the composite phase, without any O-vacancy. Let us call them $E_{ortho,perfect}$ and $E_{comp,perfect}$, respectively. The $E_{comp,perfect}$ values are obtained for different thicknesses of the m-phase. Then, we calculate $E_{comp,perfect} - E_{ortho,perfect}$ which signifies the favorability of m-phase formation at the interface (i.e., less positive this energy difference, more likely it is to form the interfacial m-phase). The red dotted line in Figure 2a shows the result. The energy difference is quite high, suggesting low possibility of m-phase dead layer formation. To understand the effect of O-vacancy, we then calculate the energy difference between composite and pure o-phases with eccentric O-vacancy ($E_{comp,ecc} - E_{ortho,ecc}$) and plot the result in Figure 2a (black dotted line). Here, the energy difference is significantly reduced compared to vacancy-free interface. Therefore, the favorability of m-phase dead layer formation greatly enhances in the presence of eccentric O-vacancy. We repeat these calculations for the centric O-vacancy. Hence, we calculate $E_{comp,cen} - E_{ortho,cen}$, i.e., the difference in the energies of the composite and pure o-phases, both having an O-vacancy in the centric layer (black dotted line in Figure 2b). The results indicate that energy difference increases when O-vacancy forms in the centric layer as compared to vacancy-free interface. This suggests that the m-phase formation is less likely to occur if the O-vacancy is in the centric layer.

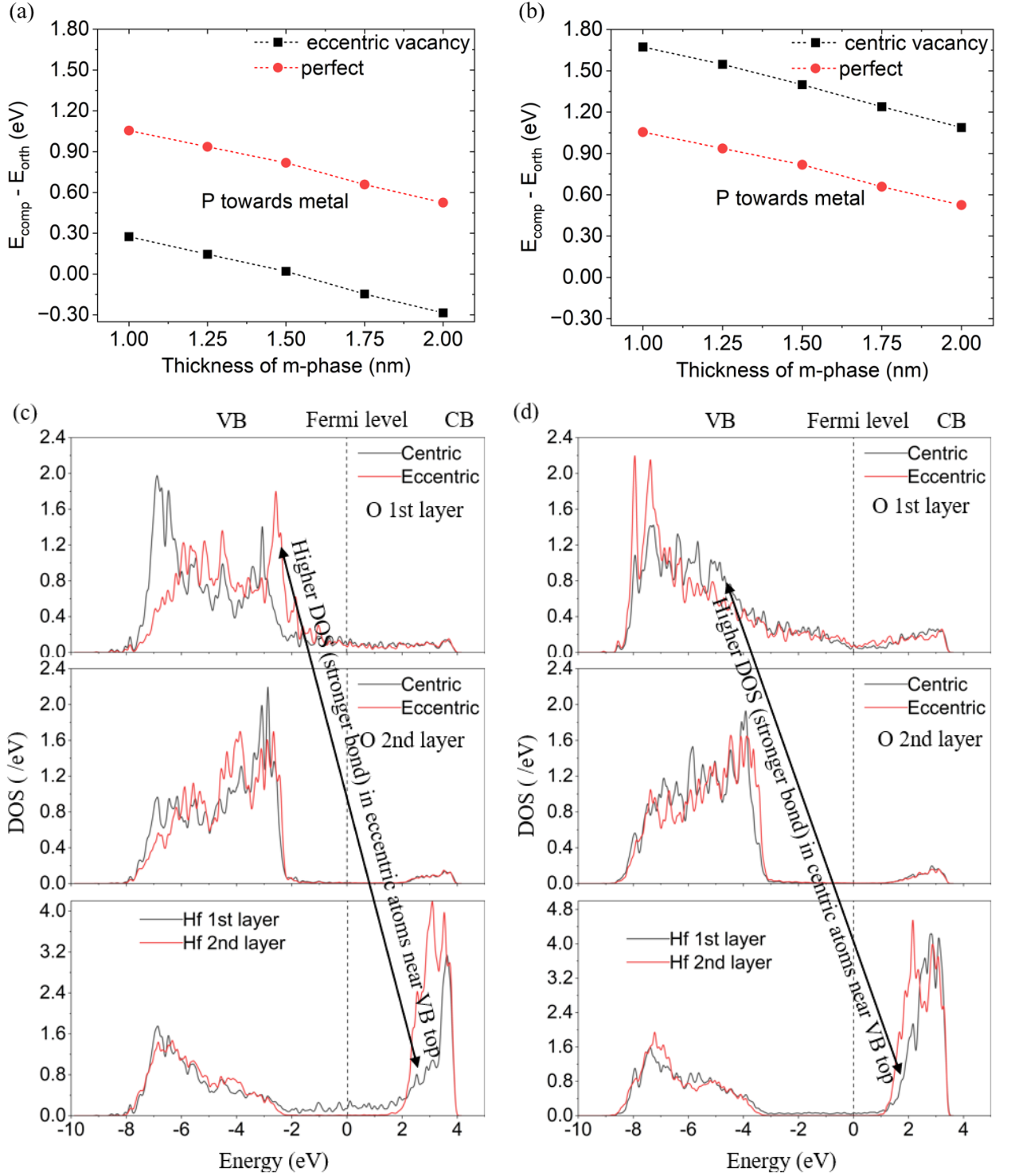

**Figure 2** Effect of oxygen vacancy at (a) eccentric and (b) centric layer on the energy difference of composite and pure o-phase with polarization pointing towards metal. O-vacancy at eccentric layer increases the likelihood of m-phase dead layer formation as opposed to the decrease for centric O-vacancy. Projected density of states (PDOS) of centric and eccentric O atoms and Hf atoms in (c) pure o-phase and (d) composite structure of 1.5 nm m-phase for first two layers. In pure o-phase, $V_{0,orth,ecc} > V_{0,orth,cen}$. With m-phase, $V_{0,comp,ecc} < V_{0,comp,cen}$ for the interfacial layer and $V_{0,comp,ecc} > V_{0,comp,cen}$ for layers away from the interface.

To understand this phenomenon, we calculate the atomic-layer-wise projected density of states (PDOS) for pure o- and composite phases without any oxygen vacancy. The Hf atoms have higher density of states in the conduction band (CB) due to their metallic nature. On the other hand, in the O atoms, the dominant density of states lies in the valence bands (VB) because of their partially filled atomic p-orbitals. The Hf-O bond strength depends on the PDOS near the top of the VB of O atoms and near the bottom of the CB of Hf atoms i.e. the higher the PDOS, the stronger is the bond. Now, if we compare the PDOS of eccentric and centric O atoms at the interface (1$^{st}$ layer) in pure o-phase (Figure 2c), eccentric O atoms have higher density of states at the top of the VB. Thus, compared to centric O atoms, eccentric O atoms can form atomic bonds with Hf atoms more easily. As a result, Hf-O bond is stronger in eccentric O atoms. Therefore, oxygen vacancy formation energy for eccentric O atoms ($V_{0,orth,ecc}$) is also larger compared to centric O atoms ($V_{0,orth,cen}$) in pure o-phase, i.e., $V_{0,orth,ecc} > V_{0,orth,cen}$ (refer to Figure 1c). Similarly, if we repeat this for the composite phase and compare the PDOS of interfacial (1st layer) eccentric O atom to centric O atom near the top of VB (Figure 2d), we observe that eccentric atom has lower density of states. As a result, Hf-O bond is weaker in eccentric O atom and hence, $V_{0,comp,ecc} < V_{0,comp,cen}$ (refer to Figure 1c). We re-emphasize that this relation is valid for only the interfacial (1st layer) oxygen atoms at site C and is opposite to the relation in bulk m-phase shown in Figure 1c. Besides the above relations, Figure 1c suggests that $V_{0,comp,ecc} < V_{0,orth,ecc}$ and $V_{0,comp,cen} > V_{0,orth,cen}$. From the definition of $V_0$ in (1), we obtain

$$(E_{comp,ecc} - E_{comp,perfect} + \frac{\mu_{O_2}}{2}) < (E_{ortho,ecc} - E_{ortho,perfect} + \frac{\mu_{O_2}}{2})$$

$$\Rightarrow E_{comp,ecc} - E_{ortho,ecc} < E_{comp,perfect} - E_{ortho,perfect} \quad (2)$$

and $\left(E_{comp,cen} - E_{comp,perfect} + \frac{\mu_{O_2}}{2}\right) > (E_{ortho,cen} - E_{ortho,perfect} + \frac{\mu_{O_2}}{2})$

$$\Rightarrow E_{comp,cen} - E_{ortho,cen} > E_{comp,perfect} - E_{ortho,perfect} \quad (3)$$

Equation (2) suggests that eccentric O-vacancy reduces the energy difference of composite phase and pure o-phase as shown in Figure 2a. Therefore, the possibility of interfacial m-phase dead layer formation increases significantly with eccentric O-vacancy. On the contrary, (3) suggests that centric O-vacancy increases the energy difference of composite phase and o-phase as shown in Figure 2b. Thus, the m-phase dead layer formation becomes energetically costlier with O-vacancy in the centric layer.

### 2.1.2 Polarization away from metal

Now, let us turn our attention to the case when polarization points away from the metal and understand the effect on the energetics of phases with O-vacancy formation. With this polarization direction, the interfaces of metal and o-phase as well as m-phase and o-phase are quite different (compared to the previous case with polarization pointing towards the metal). This is because of different bound charges and atomic arrangements in the two cases. In the pure o-phase, interfacial positive bound polarization charges appear at the eccentric layer when polarization points towards the metal. On the other hand, the interfacial bound polarization charges appearing at the eccentric layer are negative when polarization points away from the metal. Negative bound charges repel oxygen atoms due to their high electronegativity. Therefore, it is easier to form oxygen vacancy at the eccentric layer when polarization points away from the metal compared to the case when polarization points towards the metal. This results in a significant decrease in eccentric oxygen vacancy formation energy (as shown in Figure 1c) in pure o-phase. On the other hand, in composite phase, the metal-HZO interface is similar for both polarization directions. Thus, O-vacancy formation energy is similar in magnitude. As a result, although eccentric O-vacancy reduces $E_{comp} - E_{ortho}$ with polarization pointing towards the metal, it increases $E_{comp} - E_{ortho}$ with polarization pointing away from the metal.

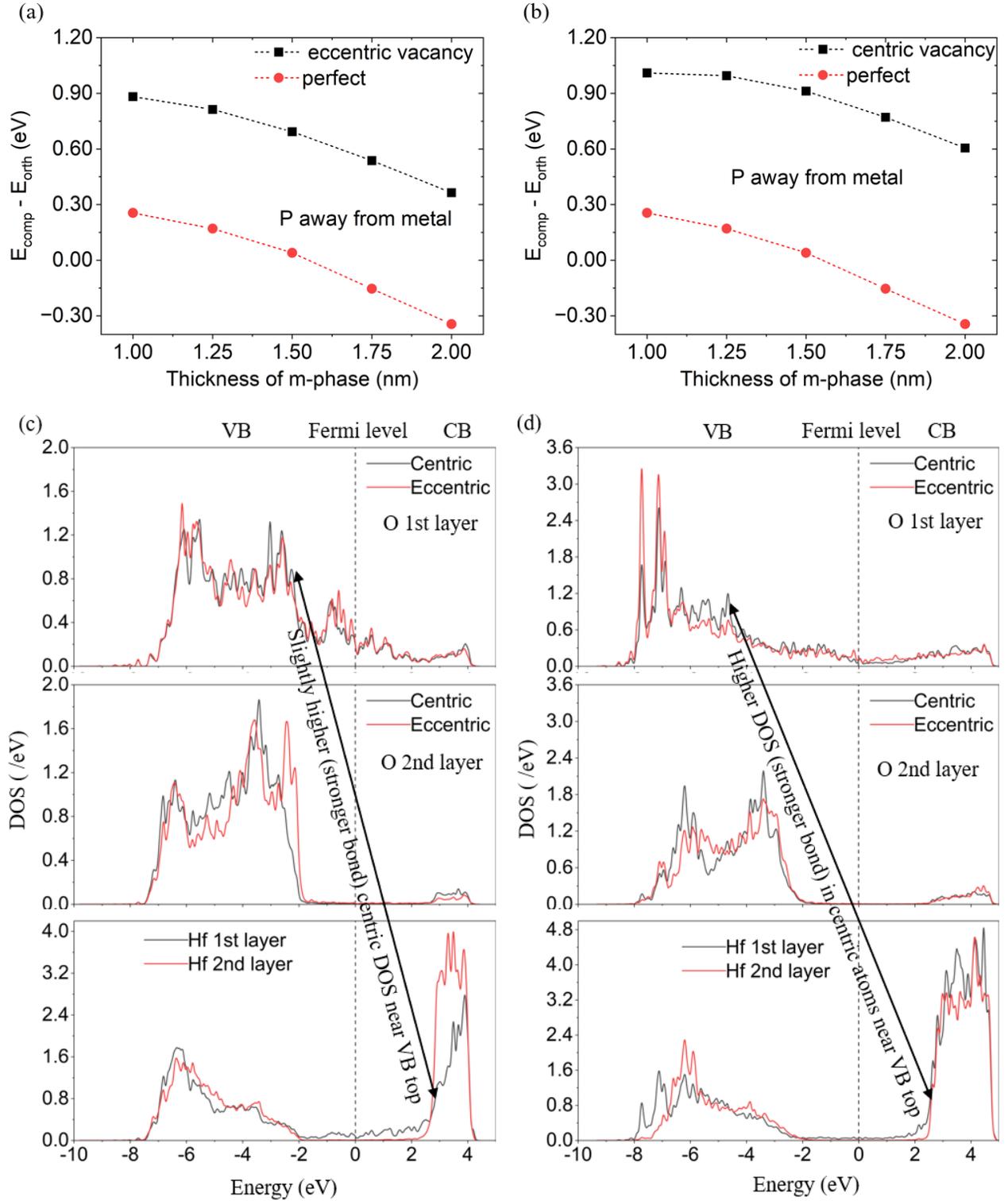

**Figure 3** Effect of oxygen vacancy at (a) eccentric and (b) centric layer on the energy difference of interfacial m-phase and o-phase with polarization pointing away from the metal. O-vacancies at both eccentric and centric layers increase the energy difference than the vacancy free interface. Thus, m-phase dead layer is less likely to form with both types of vacancies. Projected density of states (PDOS) of centric and eccentric O atoms and Hf atoms in (c) pure o-phase and (d) composite structure of 1.5 nm m-phase for first two layers. In pure o-phase, $V_{0,orth,cen}$ is slightly greater than $V_{0,orth,ecc}$. With m-phase, $V_{0,comp,ecc} < V_{0,comp,cen}$ for the interfacial layer and $V_{0,comp,ecc} > V_{0,comp,cen}$ for layers away from the interface.

The red dotted plot in Figure 3a shows the energy difference for composite and pure o-phase ($E_{comp,perfect} - E_{ortho,perfect}$) without any O-vacancy for polarization pointing away from the metal. As described in the last paragraph, in the presence of eccentric O-vacancy, the energy difference of composite and pure o-phase ($E_{comp,ecc} - E_{ortho,ecc}$) increases (black dotted plot in Figure 3a).

When the O-vacancy in the centric layer, we observe that the energy difference ($E_{comp,cen} - E_{ortho,cen}$) increases compared to the perfect interface (Fig. 3b). This trend is similar to the case of polarization pointing towards the metal. In other words, while the effect of centric O-vacancy is the same for both the polarization directions (i.e. does not favor the m-phase formation), the effect of eccentric O-vacancy is opposite (i.e. it opposes (favors) m-phase formation if the polarization points away from (towards) the metal interface).

The PDOS plots in Figure 3c-d are consistent with the abovementioned trends. From Figure 3c, for pure o-phase, we observe that PDOS is slightly higher in the centric O atom than eccentric atoms at the top of the valence band. Therefore, the oxygen vacancy formation energy at the interfacial (1$^{st}$ layer) centric atom is slightly higher than eccentric atom i.e., $V_{0,orth,cen} > V_{0,orth,ecc}$ (refer to Figure 1c). On the other hand, for composite structures, following the same argument as in the previous sub-section, Hf-O bond is the weakest when vacancy is at the position of eccentric O atom at site C (Figure 3d). Thus, $V_{0,comp,ecc} < V_{0,comp,cen}$ (refer to Figure 1c). On top of these, as Figure 1c suggests, $V_{0,comp,ecc} > V_{0,orth,ecc}$ and $V_{0,comp,cen} > V_{0,orth,cen}$. From the definition of $V_0$ in (1), we get:

$$E_{comp,ecc} - E_{ortho,ecc} > E_{comp,perfect} - E_{ortho,perfect} \qquad (4)$$

$$\text{and } E_{comp,cen} - E_{ortho,cen} > E_{comp,perfect} - E_{ortho,perfect} \qquad (5)$$

Thus, both eccentric and centric O-vacancy reduces the favorability of m-phase dead layer when polarization points away from the metal.

In summary, m-phase dead layer formation is dependent on the polarization direction and spatial distribution of oxygen vacancy at the interface. If polarization points towards the metal, eccentric O-vacancy favors the m-phase dead layer formation, but centric O-vacancy does not favor it. However, if polarization points away from the metal, m-phase dead layer is not favored energetically, regardless of the spatial position of the O-vacancy.

With this understanding of the effect of interfacial oxygen vacancy on m-phase dead layer formation, we analyze other important aspects, i.e., effect of non-interfacial O-vacancy, Zr concentration, metal electrode selection, etc. For the subsequent analysis, we will consider the case of polarization pointing towards the metal, as m-phase dead layer is only favorable for that case.

### 2.2 Oxygen Vacancies Away from the Interface

The effect of O-vacancies away from the interface along the thickness of HfO$_2$ is depicted in Figure 4. Let us first discuss the case of centric layer O-vacancy shown in Figure 4a. As the centric O-vacancy shifts from the interface to bulk, the O-vacancy formation energy increases for both the pure o- and composite phases. Initially, $V_o$ for the composite phase increases more than the pure o-phase, thus yielding a slight increase in the energy difference between the composite and pure o-phases. After that the energy difference reduces for deeper oxygen vacancies. However, for all the values of O-vacancy depths considered, the energy difference ($E_{comp,cen} - E_{ortho,cen}$) is significantly positive in magnitude. This implies that pure o-phase is the dominant phase regardless of the location of the centric O-vacancy along the thickness.

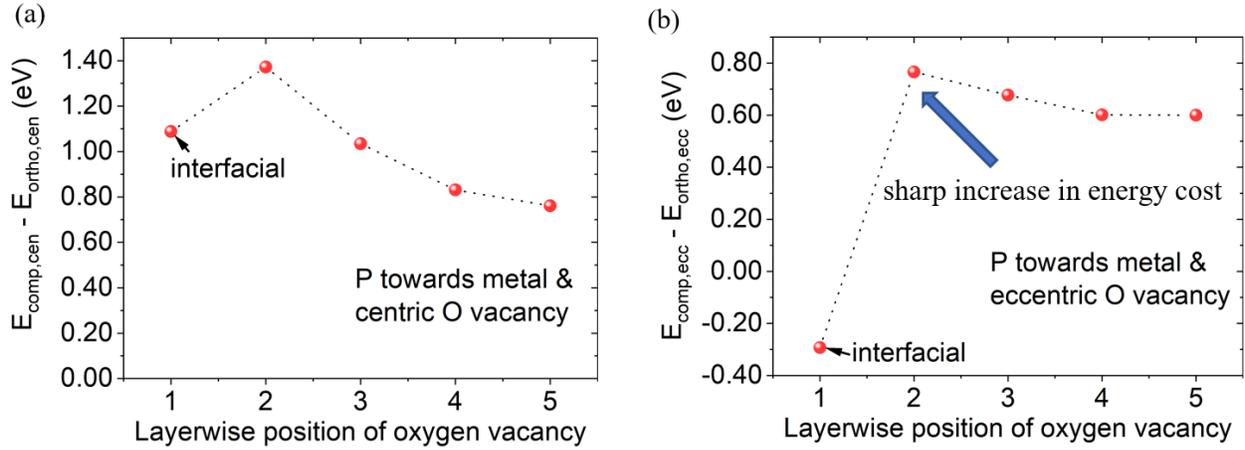

**Figure 4** Energy difference of composite and pure o-structure with respect to the depth of O vacancy for 2 nm m-phase thickness with (a) centric and (b) eccentric oxygen vacancy when polarization points towards metal. With centric O vacancy, the energy difference is still significantly high. When the eccentric O vacancy shifts away from the interface, its formation energy increases drastically as supported by Fig. 2d. Thus, m-phase formation becomes less likely when eccentric vacancy shifts from interface to bulk.

Now, let us look at the eccentric O-vacancy. In this case, the initial jump of the energy difference of composite and pure o-phase is more prominent. Figure 2d suggests that composite phase PDOS in eccentric O atoms at the 2nd layer are higher compared to centric O atoms around the top of the VB (opposite to the case of O atoms at the 1st layer). Due to this abrupt change of the PDOS characteristics from 1st to 2nd layer of composite phase, the O-vacancy formation energy at the eccentric layer increases drastically when O-vacancy moves away from the interface towards bulk. Then, $V_{0,comp,ecc}$ becomes greater than $V_{0,comp,cen}$ (similar to what we observed in Figure 1c for the bulk $HfO_2$). This results in a sudden jump in the energy difference ($E_{comp,ecc} - E_{ortho,ecc}$) from a negative to a positive value as shown in Fig. 4b. Thus, the pure o-phase becomes more favorable. This suggests that m-phase dead layer formation becomes energetically unfavorable when the oxygen vacancies are not at the interface. In summary, eccentric oxygen vacancies need to be at the interface in order to form m-phase dead layer when polarization points towards the metal.

The analysis above also hints at the possibilities of the role of eccentric O vacancies in the wake-up effect in HZO. It is well-established that during electric field cycling, oxygen vacancies redistribute within the FE sample [32] [33]. According to our discussion above, the shift of O-vacancy from the interface towards bulk increases the favorability of o-phase. This is consistent with the experimental finding of monoclinic to orthorhombic phase transition during electric field cycling as a possible mechanism of wake-up effect in HZO [34]. However, to validate the implication of O vacancy movement, further analysis of the effect of electric field on oxygen vacancy movement is needed which is out of the scope of the current work.

### 2.3 Polarization profile

We will now discuss the implication of the spatial distribution of oxygen vacancy on ferroelectric polarization. As discussed above, for centric O vacancy at the interface, pure o-phase is energetically favorable than the composite phase. On the other hand, the m-phase dead layer can form with eccentric O vacancy. Therefore, we calculate the layer wise polarization profile for pure o-phase with centric O vacancy and composite phase with eccentric O vacancy. These are illustrated in Figure 5(a) and 5(b) respectively. In the pure o-phase, polarization in all the layers except those close to the interfacial layer is equal to the bulk polarization. Thus, with vacancy at centric O atoms, the dead layer is very thin (around 0.50 nm) and is due to the interfacial relaxation only. On the other hand, in the composite phase, the m-phase region shows polarization close to zero. Thus, with interface vacancy at eccentric O atoms, dead layer is much thicker, i.e., the combined thickness of interfacial layer, m-phase, and phase boundary between m- and o-phases.

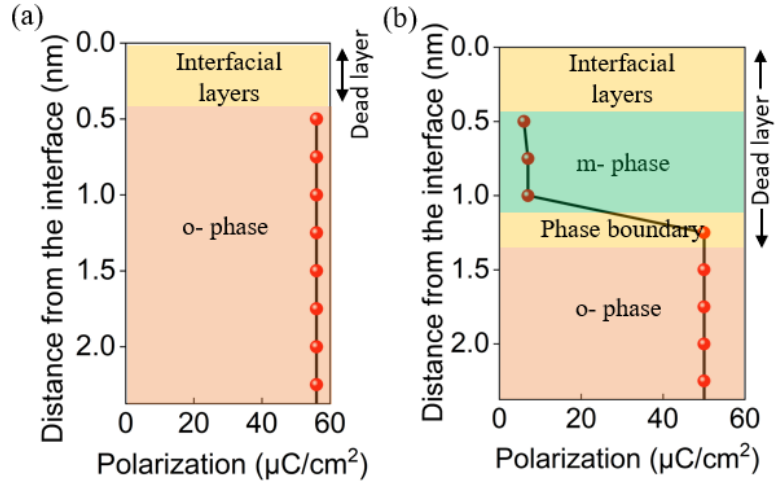

**Figure 5** Polarization profile in (a) pure o-phase with centric O vacancy and (b) composite phase with eccentric O vacancy. When O vacancy is in the centric layer, dead layer forms only due to the interfacial atomic relaxation. When vacancy is in the eccentric layer, dead layer is thicker due to m-phase formation.

### 2.4 Effect of Zr doping

So far, our discussion has been focused on undoped $HfO_2$. Here, we analyze the effect of Zr doping on m-phase dead layer formation by considering 50% Zr doped $HfO_2$ (HZO). The unit cells of o-phase and m-phase of HZO used in this calculation are shown in the left panel of Figure 6. It is well known that as Zr is added in HfO$_2$, the o-phase becomes more energetically favorable compared to the m-phase [35]. Earlier we have showed that for centric O-vacancy, energy difference between composite and pure o-phase is large. We observe that Zr doping further increases the energy difference between the composite and pure-o phases and makes m-phase dead layer even less likely. Now, recall, for the eccentric interfacial O-vacancy, m-phase dead layer can form for undoped HfO$_2$. However, for HZO, the energy difference between the composite and pure-o phases increases as shown in Figure 6, thus reducing the likelihood of the formation of the composite phase. These results confirm that with increasing Zr doping, the dead layer formation due to m-phase becomes less likely.

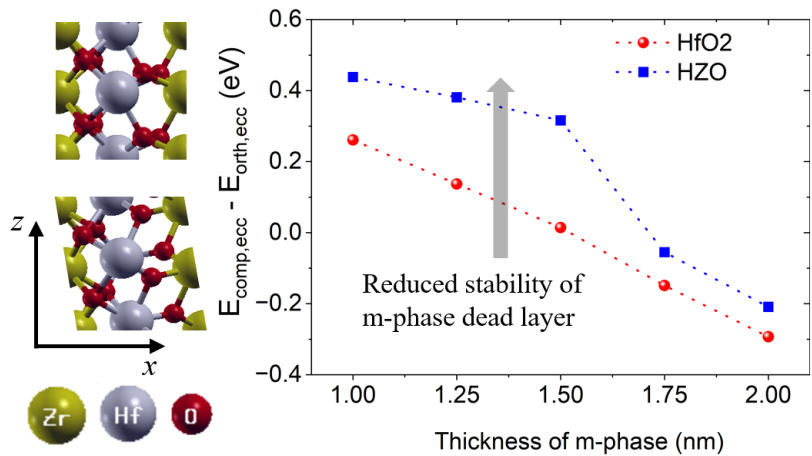

**Figure 6** Energy difference of composite and pure o-structure of $HfO_2$ and HZO (unit cells attached) with O-vacancy in the eccentric layer for polarization towards metal. Zr doping slightly increases m-phase energy and thus reduces the favorability of m-phase dead layer formation compared to $HfO_2$.

## 2.5 Effect of metal electrodes

Until now, we have only considered Tungsten (*W)* as the metal electrode. To consider the effect of metal electrode, we now analyze another widely used metal '*Pt*'. Our results in Fig. 7a indicate that with Pt metal electrode, the energy difference between composite and pure o-phase is always positive, i.e., significantly larger than that for W. Thus, the dead layer formation due to interfacial m-phase is less likely with Pt. Here, the dead layer is predominantly due to the interfacial relaxation of o-phase and the thickness of dead layer is around 1 unit cell, (like for the case of centric O-vacancy for W electrode). The fact that dead layer thickness is less for Pt electrode compared to W electrode agrees with the experimental results shown in Reference [18].

To understand this trend further, we discuss different factors that affect the relative energy shift due to the metal electrode. These include induced strain with HfO$_2$ and work function of metal. Compressive strain reduces the o-phase energy compared to the m-phase [21]. Thus, the favorability of m-phase dead layer formation reduces with higher compressive strain /less tensile strain. To understand the effect of metal work function, let us assume that the Fermi level of metal lies inside the bandgap of HfO$_2$, as will be the case with all the metals discussed in this work. With larger work function metal, Fermi level lies closer to the VB of O atoms. Thus, the bond strength between the metal electrode and O atoms at the interface increases, i.e., O-vacancy formation energy is increased. However, the VB edge of the m-phase is lower in energy (i.e. further away from the metal Fermi level) compared to the o-phase, as can be observed from Figure 2c and 2d. Therefore, the relative increase in the strength of the O-metal bond (i.e., O-vacancy formation energy) due to work function increase is larger in the o-phase ($V_{0,orth}$) compared to that in the m-phase ($V_{0,comp}$). Hence, as the work function increases, $V_{0,comp} - V_{0,orth}$ decreases and the favorability of m-phase dead layer formation increases.

The relaxed lattice parameters of bulk HfO$_2$ and metals indicate that strain induced in HfO$_2$ is more compressive for Pt compared to W [36]. On the other hand, Pt has higher work function compared to W. Thus, the effect of induced strain and work function on m-phase stability is opposite to each other. As Pt shows higher energy difference between composite and pure o-phase, it suggests that between W and Pt, the strain has a larger impact on the relative energy compared to the work function.

We also analyze the effect of using other noble metals (Pd, Rh, Os and Ru) on the likelihood of m-phase dead layer formation. The dependence of the energy difference for each of these metals (in addition to Pt and W) with respect to corresponding strain and work function values are shown in Fig. 7b and 7c, respectively. There is no regular trend in the graphs because of the complex interplay among strain and work function. However, all the 5 metals except W show high energy differences between the composite and pure o-phase, with Pd and Pt being the highest. This suggests that m-phase dead layer formation is less likely with Pt, Pd, Rh, Os and Ru electrode, Thus, with these electrodes, a thin dead layer is expected mainly due to the interface relaxation with o-phase.

From the analysis of the effect of metal electrodes, we can infer that Pt and Pd are the most suitable metals for minimizing dead layer thickness. Thus, they are conducive to obtaining larger memory windows. However, for optimization of the memory functionalities, other factors, such as depolarization field and band offset are also important. For that purpose, thorough investigation of each of these factors is necessary in addition to the dead layer effect presented in this work.

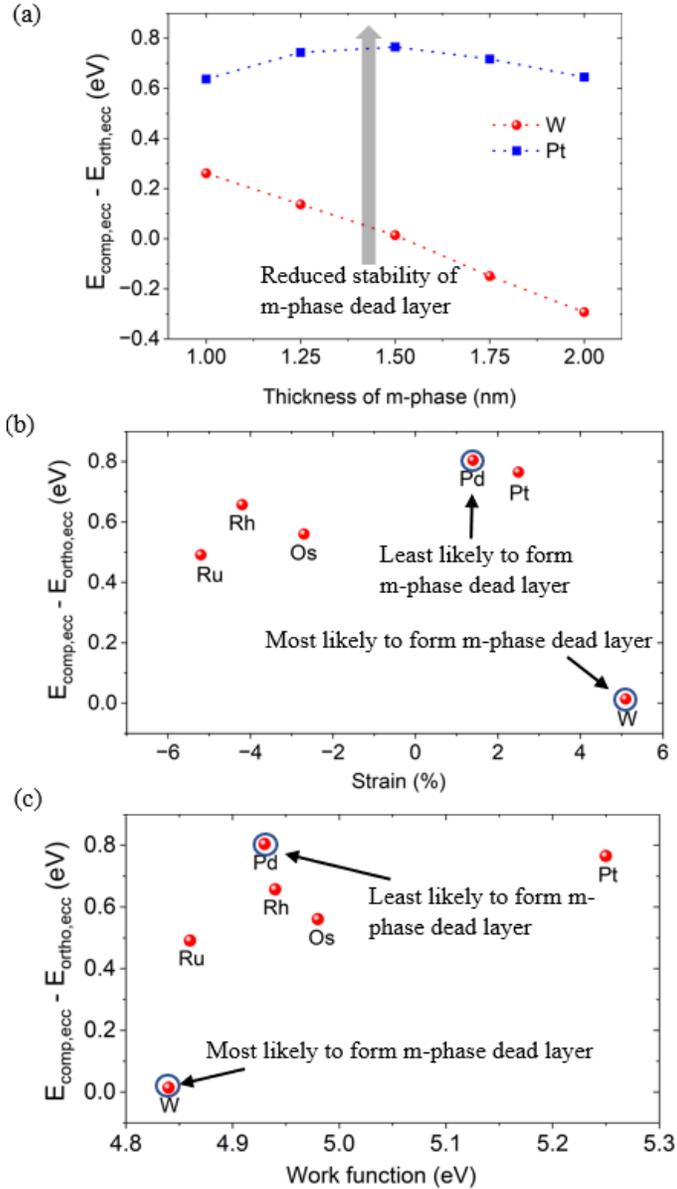

**Figure 7** (a) Energy difference of composite and pure o-structure for (a) Tungsten (W) and Platinum (Pt) metal electrode with O-vacancy in eccentric layer and polarization towards metal. In Pt, the energy difference is always highly positive; thus, the possibility of m-phase dead layer formation is less. Energy difference of composite and pure o-structure for different metals with 1.5 nm m-phase in $HfO_2$ as a function of (b) induced strain in metal and (c) work function of metal. Among the calculated metals, Pd and W has the least and most likelihood of m-phase dead layer formation, respectively.

### 3. Conclusion

We performed first principles density functional theory (DFT) calculations to predict a possible mechanism of dead layer formation due to monoclinic phase in $HfO_2$-based ferroelectrics. We find that for W metal electrode, when polarization points towards the metal, interfacial vacancies in trigonally bonded oxygen sites (eccentric) make the formation of interfacial m-phase energetically comparable to pure o-phase. This interfacial m-phase forms a thick dead layer. On the other hand, vacancies in tetrahedral sites (centric) of oxygen atoms make interfacial m-phase formation energetically unfavorable. For polarization pointing away from the metal or vacancies migrated away from the interface, interfacial m-phase dead layer is less favorable, regardless of the site of oxygen vacancy. Therefore, for these cases, only the interfacial o-phase forms a thin dead layer. The probability of such m-phase dead layer formation is reduced with Zr doping, as well. Further, our analysis shows that noble metals such as Pt and Pd metal electrodes have the least possibility of m-phase dead layer formation. Therefore, this work not only provides insights into O-vacancy-dependent dead layer formation mechanisms but also offers a guideline for metal choice targeting the optimization of dead layer in fluorite ferroelectric devices.

## Methodology:

We use the first principles density functional theory (DFT) in Quantum Espresso software package [37]. Atomic structures are generated using XCrysden software package [38]. We employ Projector-Augmented Wave (PAW) pseudopotentials with non-linear core correction and Perdew–Burke–Ernzerhof generalized gradient approximation (GGA-PBE) as exchange correlation functional. Polar orthorhombic $Pca2_1$ phase (o-phase) and non-polar monoclinic $P2_1/c$ phases (m-phase) are optimized with a maximum of $10^{-6}$ Ry error in ionic minimization energy, $10^{-4}$ Ry/Bohr force tolerance, 60 Ry of kinetic energy cut-off and 6×6×6 Monkhorst-pack grid of k points. Cubic $Fm\bar{3}m$ phases of metals are relaxed, cleaved along the (111) miller indices and strained in the in-plane directions to lattice match with $HfO_2$. 5 layers of (2×2) supercell of the cleaved surface are put on top of (1×1) surface of desired $HfO_2$ to form the heterostructure. Vacuum of 10 A° on both sides of the heterostructure is formed giving a total of 20 A° gap. The validity of the assumed vacuum thickness is verified by energy convergence with respect to vacuum layer thickness. As we are interested in dead layer formation with FE phase, we fix the atomic configuration in the outermost layer of the FE corresponding to the bulk values. Then, we atomically relax the heterostructures with fixed in-plane lattice parameters equal to the bulk polar phase. For these heterostructure calculation, we use 4×4×1 Monkhorst-pack grid of *k* points and Gaussian smearing of 0.01 Ry. For MFM calculation, 2×2×1 Monkhorst-pack grid of *k* points and Gaussian smearing of 0.01 Ry are used. Note, at zero K temperature and unstrained conditions, tetragonal phase with metal is unstable and hence, we only consider m-phase as the non-polar phase. Polarization is calculated according to the modern theory of polarization [39]. The Projected Density of States (PDOS) is obtained from data postprocessing packages in Quantum Espresso and used in MATLAB to generate layer wise plots. Due to the widely known bandgap underestimation problem with GGA functional, the absolute numerical values from band calculation may not be exact. But they are useful to understand the relative trends.

# Supplementary Information
# Oxygen Vacancy-Induced Monoclinic Dead Layers in Ferroelectric Hf$_x$Zr$_{1-x}$O$_2$ With Metal Electrodes

Tanmoy Kumar Paul[1], Atanu Kumar Saha[1], and Sumeet Kumar Gupta[1]
[1]Purdue University, West Lafayette, Indiana, USA, email: paul115@purdue.edu

| | U | | | | D | | | | $m$-phase | | |
|---|---|---|---|---|---|---|---|---|---|---|---|
| Hf | 0.484 | 0.466 | 0.000 | Hf | 0.484 | 0.467 | 0.000 | Hf | 0.475 | 0.457 | 0.042 |
| Hf | 0.016 | -0.034 | 0.000 | Hf | 0.016 | -0.033 | 0.000 | Hf | 0.025 | -0.043 | -0.042 |
| O | 0.816 | 0.634 | 0.144 | O | 0.816 | 0.865 | 0.355 | O | 0.816 | 0.673 | 0.099 |
| O | 0.684 | 0.134 | 0.144 | O | 0.684 | 0.365 | 0.355 | O | 0.684 | 0.327 | 0.401 |
| O | 0.213 | 0.271 | 0.252 | O | 0.214 | 0.229 | 0.248 | O | 0.199 | 0.243 | 0.227 |
| O | 0.287 | 0.771 | 0.252 | O | 0.286 | 0.729 | 0.248 | O | 0.301 | 0.757 | 0.273 |
| Hf | 0.016 | 0.532 | 0.500 | Hf | 0.016 | 0.534 | 0.500 | Hf | 0.025 | 0.543 | 0.458 |
| Hf | 0.484 | 0.032 | 0.500 | Hf | 0.484 | 0.034 | 0.500 | Hf | 0.475 | 0.043 | 0.542 |
| O | 0.684 | 0.364 | 0.644 | O | 0.684 | 0.136 | 0.855 | O | 0.684 | 0.173 | 0.901 |
| O | 0.816 | 0.864 | 0.644 | O | 0.816 | 0.636 | 0.855 | O | 0.816 | 0.827 | 0.599 |
| O | 0.287 | 0.728 | 0.752 | O | 0.286 | 0.772 | 0.748 | O | 0.301 | 0.743 | 0.773 |
| O | 0.213 | 0.228 | 0.752 | O | 0.214 | 0.272 | 0.748 | O | 0.199 | 0.257 | 0.727 |

**Figure S1** Fractional atomic coordinates of upward (U) and downward (D) polarized unit cells of o-phase and m-phase considered in this work.

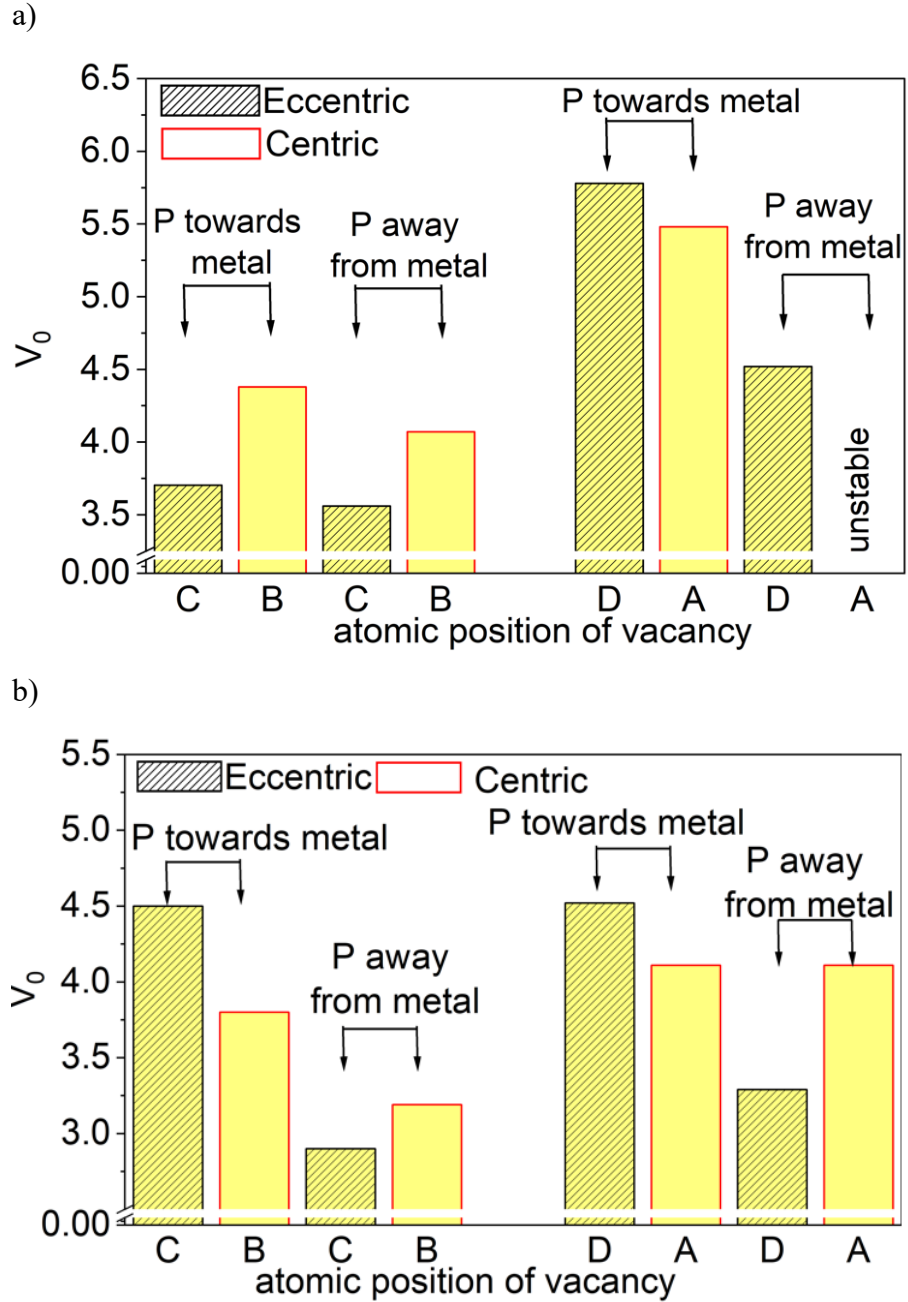

**Figure S2** Oxygen vacancy formation energy ($V_0$) of all centric and eccentric interfacial O atoms in a) composite and b) pure o-phase shown in Figure 1b. Eccentric *C* atom demonstrates lower $V_0$ than eccentric *D* atom and Centric *B* atom gives lower $V_0$ than centric *A* atom, irrespective of the polarization direction. Therefore, vacancies at B and C atomic positions are taken as the default centric and eccentric O vacancies, respectively.

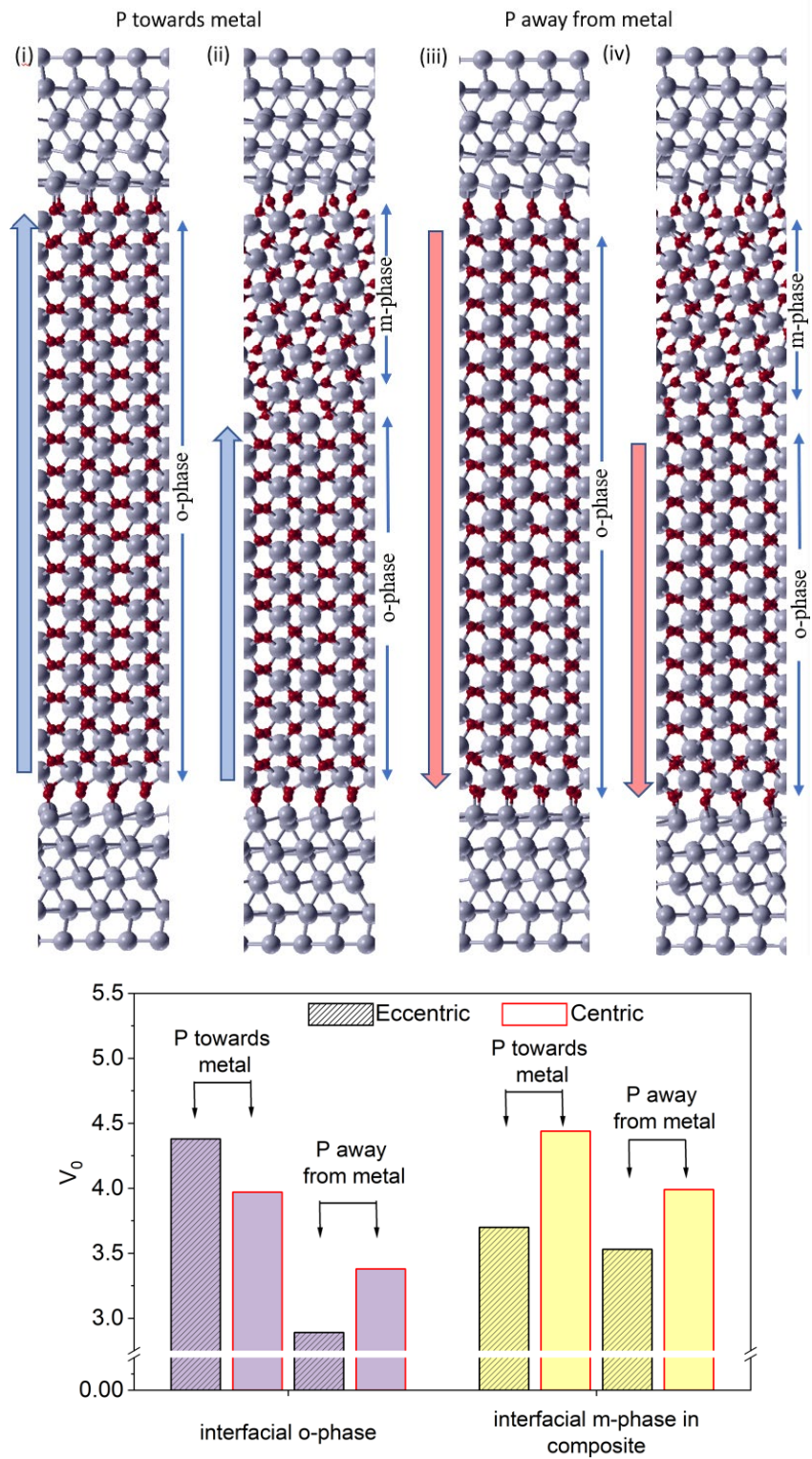

**Figure S3** a) Metal-HfO$_2$-metal structures of pure o-phase and composite phase with 1.5 nm m-phase dead layer at the top interface for both polarization directions. b) Corresponding interfacial O vacancy formation energies. The oxygen vacancy formation energies are consistent with metal-FE slab results shown in Figure 1c.